\documentclass[10pt,twocolumn,reqno,amsmath,amssymb,aps,prb]{revtex4-2}
\usepackage[utf8]{inputenc}

\DeclareUnicodeCharacter{00A0}{~}  % Non-breaking space

\usepackage{graphicx}
\usepackage[justification=justified, singlelinecheck=false]{caption}
\usepackage[justification=justified]{subcaption}

\usepackage{hyperref}    
\usepackage{xcolor}
\usepackage{leftindex}

\usepackage{dcolumn}% Align table columns on decimal point
\usepackage{bm}% bold math
\begin{document}

\preprint{APS/123-QED}

\title{Dipolar-exchange spin waves in thin bilayers}%

\author{Rob den Teuling$^1$}

\author{Ritesh Das$^1$}
\author{Artem V. Bondarenko$^{1,2}$}

\author{\\ Elena V. Tartakovskaya$^{1,2,3}$}
\altaffiliation[Corresponding author:]{ olena.tartakivska@gmail.com.}

\author{Gerrit E. W. Bauer$^{4,5}$}

\author{Yaroslav M. Blanter$^1$ \vspace{3mm}}

\affiliation{${}^1$ \hspace{-3mm} Kavli Institute of Nanoscience, Delft University of Technology, Lorentzweg 1, 2628 CJ, Delft, The Netherlands, \vspace{2mm}}%

\affiliation{${}^2$ \hspace{-3mm} V.G. Bar'yakhtar Institute of Magnetism of the NAS of Ukraine, 36b Vernadsky Boulevard, 03142 Kiev, Ukraine,\hspace{-3mm} \vspace{2mm}}

\affiliation{${}^3$ \hspace{-3mm} Institute of Spintronics and Quantum Information, Faculty of Physics and Astronomy, Adam Mickiewicz University, Poznań, Uniwersytetu Poznańskiego 2, 61-614 Poznań, Poland,\hspace{-2.5mm} \vspace{2mm}}

\affiliation{${}^4$ \hspace{-3mm} WPI-AIMR \& IMR  \& CSIS, Tohoku University, 2-1-1 Katahira, Sendai 980-8577, Japan,
\vspace{2mm}}

\affiliation{${}^5$ \hspace{-3mm} Kavli Institute for Theoretical Sciences, University of the Chinese Academy of Sciences, Beijing 10090, China. \vspace{2mm}}

\date{\today}% It is always \today, today,

\begin{abstract}
We investigate the dipolar-exchange spin wave spectrum in thin ferromagnetic bilayers with in-plane magnetization, incorporating interlayer exchange coupling and intra- and interlayer dipolar interactions. In the continuum approximation we analyze the nonreciprocity of propagating magnetic stray fields emitted by spin waves as a function of the relative orientation of the layer magnetizations that are observable by  magnetometry of synthetic antiferromagnets or weakly coupled type-A van der Waals antiferromagnetic bilayers as a function of an applied magnetic field. 
\end{abstract}

\maketitle

%\tableofcontents

\section{\label{sec:level1}Introduction}

Since the scientific legacy of Victor Bar'yakhtar is vast and multifaceted, a comprehensive overview falls outside the scope of this article. Here, we focus on a select subset of his contributions that strongly influences the scientific perspective of one of the authors (E.V.T.), who is directly associated with Bar'yakhtar’s research school. The methods developed by Bar'yakhtar and Maleev to describe neutron scattering by magnetic materials \cite{osti_4713671,Maleev2002PolarizedNS} enabled the understanding of a broad array of experiments, such as the spin wave spectra of multilayered rare-earth metal systems \cite{grunwald,PhysRevB.82.014426}, the ground-state magnetic configurations and phase transitions observed in neutron scattering experiments on segmented nanowire arrays \cite{10.1021/acsnano.7b03488,inbook}, and the behavior of thin layers with itinerant ferromagnetic phases \cite{7,8}. The work of Bar'yakhtar and his collaborators on magnetic soliton dynamics \cite{9,10,11} laid the foundation for burgeoning field that investigates three-dimensional magnetic textures \cite{12,13}.

Here we address the problem of exchange-dipole spin waves (SWs) in planar ferromagnetic (FM) bilayers with in-plane magnetization. We consider both the dipole and exchange interactions. For previous studies of this problem see \cite{14,15}, and references therein. We approach it without a series expansion in \(ka\), where \(k\) is the modulus of the wave vector and \(a\) is a thickness of the layer. We employ the continuum approximation, based on the seminal work by Akhiezer, Bar’yakhtar, and Peletminsky \cite{Akhiezer1959} (see also monographs and reviews \cite{16,MagWav}) and discuss its applicability to atomically thin ferromagnetic layers.

We start by calculating the dipolar fields and SW frequencies in a FM layer that is thinner than its exchange length such that the low-frequency excitations are constant normal to the layer.  In permalloy, for example, the thickness should not exceed 30 nm \cite{17}. The impact of various boundary conditions on the magnetization profile across the thickness and the corresponding frequencies was analyzed in \cite{18}.

\section{\label{sec:level1}Theoretical framework}

\begin{figure} [htp]

\includegraphics[clip,width=0.7\columnwidth]{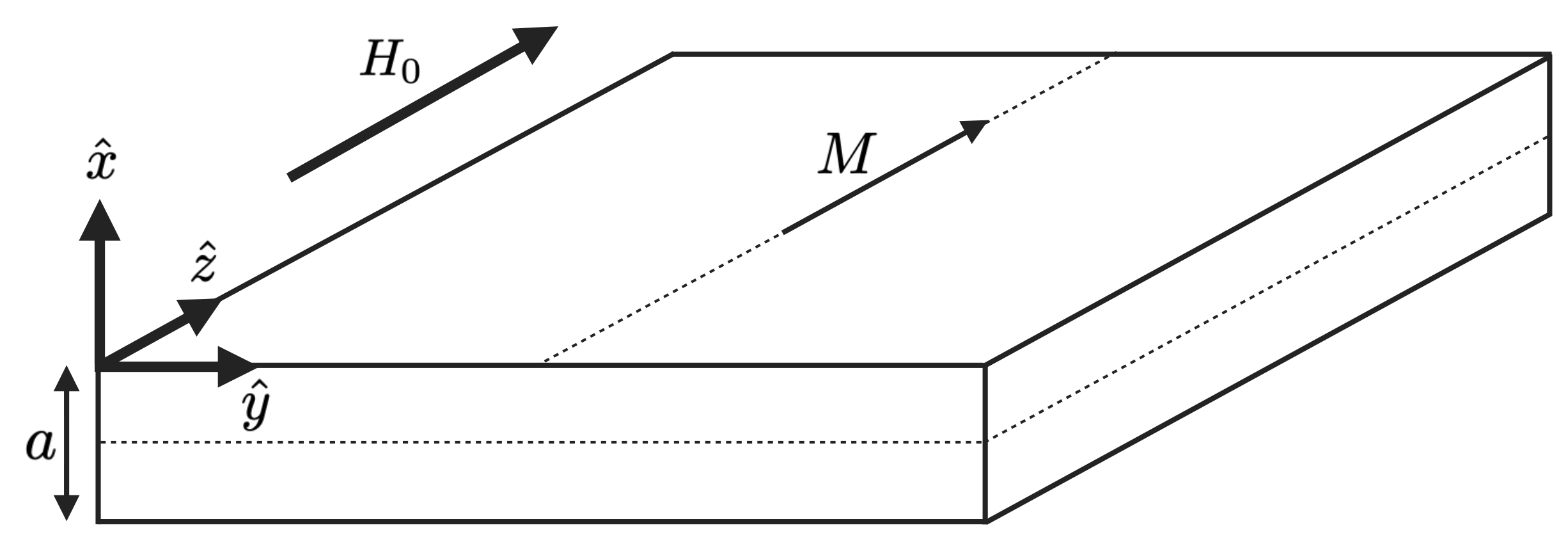}\label{1a}%

\caption{ \small Ferromagnetic layer with  external magnetic field along the in-plane magnetization.}
\label{fig1}
\end{figure}

\hyperref[1a]{Fig. 1} sketches a ferromagnetic layer of thickness \(a\) in the \((y,z)\)-plane. The magnetization \(M_0\) and the external magnetic field \(\Vec{H}\) both lie along the \(z\)-axis. For weak excitations
\begin{equation} \label{1}
    \begin{aligned}
    \Vec{M} & = \Vec{M}_0 + \Vec{m}(y,z)e^{i\Omega t} \\
    & = [m_x(y,z)e^{i\Omega t} , m_y(y,z)e^{i\Omega t} , M_0]^T \\
    \end{aligned}
\end{equation}

\noindent
The linearized Landau-Lifshitz (LL) equation takes the form
\begin{equation} \label{2}
    \begin{aligned}
        i\frac{\Omega}{\mu_0 \gamma} m_x & = (H_0 - M_0 D \Delta)m_y + M_0 \frac{\partial \Phi}{\partial y}, \\
        -i\frac{\Omega}{\mu_0 \gamma} m_y & = (H_0 - M_0 D \Delta)m_x + M_0 \frac{\partial \Phi}{\partial x}. \\
    \end{aligned}
\end{equation}

\noindent
Here \(\gamma/2\pi = 29.6\) GHz/T is the gyromagnetic ratio, \(H_0= |\Vec{H}|\), \(M_0=|\Vec{M_0}|\), \(D=(2A)/(\mu_0 M_0^2)\) is the squared exchange length \cite{16}, \(A\) is the exchange stiffness, and the magnetic scalar potential \(\Phi\) satisfies the magnetostatic equation
\begin{equation} \label{3}
    \begin{aligned}
        \nabla^2 \Phi & = \Vec{\nabla}\cdot\Vec{M}, \hspace{5mm} -\frac{a}{2} \leq x \leq \frac{a}{2}, \\
        \nabla^2 \Phi & = 0, \hspace{5mm} x<-\frac{a}{2}, \hspace{5mm} x > \frac{a}{2}.
    \end{aligned}
\end{equation}

\noindent
In addition, the boundary conditions for the magnetic potential must be satisfied, corresponding to continuity of the tangential component of the magnetic field vector and the normal component of the magnetic induction vector. 
	
Two methods can be used for the analytical evaluation of the dipole-exchange SW spectrum. One approach, proposed by De Wames and Wolfram \cite{19}, involves solving \hyperref[2]{Eqs. (2)} and \hyperref[3]{(3)} simultaneously as a system of differential equations with corresponding boundary conditions using a trial set of eigenfunctions. However, as it turned out, this approach is not suitable for a broad range of sample geometries and magnetic moment directions. In fact, its effectiveness is limited to the cases where external field and the saturation magnetization are completely in-plane or perpendicular to the surface in infinite layers, as well as in infinite wires where magnetic moments align along the wire axis \cite{20,21}. Generally speaking, this method gives valid results only if an exact solution exists, but for dipolar-exchange SW problems this is quite a rare occurrence since the exchange and dipolar operators usually have different eigenfunctions; see, however, Ref. \cite{eigen}.

For the general case an alternative approach was proposed \cite{16,16-2}, where \hyperref[3]{Eq. (3)} is solved separately using
\begin{equation} \label{4}
    \Phi(\Vec{r}) = \frac{1}{4\pi}\int d\Vec{r}{\hspace{1mm}'}\left(\Vec{m}{'}\cdot\Vec{\nabla}{'}\frac{1}{|\Vec{r}-\Vec{r}{\hspace{1mm}'}|^2}\right)
\end{equation}

\noindent
where the integration over \(r{'}\) is over the volume of the magnetic material, and \(m{'}(r{'})\) should be computed self-consistently with \hyperref[2]{Eq. (2)}.  In such a case the boundary conditions for the magnetic potential are satisfied automatically. \hyperref[4]{Eq. (4)} holds both inside and outside the magnetic material. If an exact solution does not exist, an approximate one can be obtained by perturbation theory. This method has resolved the majority of spin dynamics problems not only in layers, but also in confined magnetic elements under different magnetic field configurations, and will be used in the following sections. For an extended layer we chose the plane wave Ansatz
\begin{equation} \label{5}
    \begin{aligned}
        m_x(y,z) & = Ae^{ik_yy+ik_zz} = Ae^{i \Vec{k}\cdot \Vec{\rho}{\hspace{0.5mm}}}, \\
        m_y(y,z) & = Be^{ik_yy+ik_zz} = Be^{i \Vec{k}\cdot \Vec{\rho}{\hspace{0.5mm}}},
    \end{aligned}
\end{equation}

\noindent
where \(\Vec{k}=(0,k_y  ,k_z)\) is the wave vector and \(\Vec{\rho}=(0,\rho_y  ,\rho_z)\). Inserting \hyperref[5]{Eq. (5)} into \hyperref[4]{Eq. (4)} gives
\begin{equation} \label{6}
    \begin{aligned}
        \Phi=\int dV^{'} & \Big(m_x (y{'},z{'})  \frac{\partial}{\partial x{'}}  \frac{1}{|\Vec{r}-\Vec{r}{\hspace{1mm}'}|} \\
        & + m_y (y{'},z{'})  \frac{\partial}{\partial y{'}}  \frac{1}{|\Vec{r}-\Vec{r}{\hspace{1mm}'}|}\Big)   
    \end{aligned}
\end{equation}

\noindent
Next we show that the plane waves are functions of the dipolar operator as well. For this purpose we use the Fourier representation
\begin{equation} \label{7}
    \frac{1}{|\Vec{r}-\Vec{r}{\hspace{1mm}'}|} = \int \frac{d\Vec{q}}{(2\pi)^2} \frac{2\pi}{q} e^{i\Vec{q}\cdot(\Vec{\rho}-\Vec{\rho}{'})} e^{-q|x-x{'}|},
\end{equation}

\noindent
where \(\Vec{q}=(0,q_y ,q_z)\) and \(q=|\Vec{q}|\). The integrals run from \(-\infty\) to \(\infty\). The Fourier components of the integrals in \hyperref[6]{Eq. (6)} are then separable and read inside the layer (\(-a/2<x<a/2\))
\begin{equation} \label{8}
    \begin{aligned}
        \Phi_{in} = e^{i\Vec{k}\cdot\Vec{\rho}}&\Big( A \frac{1}{2k} e^{-ak/2} (e^{kx} - e^{-kx}) \\
        & - B i \frac{ik_y}{2k}\big(\frac{2}{k} - \frac{e^{-\frac{ak}{2}(e^{kx} - e^{-kx})}}{k}\big) \Big),
    \end{aligned}
\end{equation}

\noindent
where \(k=\sqrt{k_x^2+k_y^2}\). Calculating the dipolar magnetic fields averaged over the thickness
\begin{equation} \label{9}
    \Vec{h} = -\int^{\frac{a}{2}}_{-\frac{a}{2}} dx \Vec{\nabla}\Phi,
\end{equation}

\noindent
we obtain
\begin{equation} \label{10}
    \begin{aligned}
        h_x(\Vec{r}) & = -A e^{i\Vec{k}\cdot\Vec{\rho}}f(k), \\
        h_y(\Vec{r}) & = -B e^{i\Vec{k}\cdot\Vec{\rho}}\frac{k_y^2}{k^2}(1-f(k)), \\
    \end{aligned}
\end{equation}

\noindent
where
\begin{equation} \label{11}
    f(k) = \frac{1-e^{-ak}}{ak}.
\end{equation}

\noindent
Substituting  \hyperref[5]{Eqs. (5)} and \hyperref[10]{(10)} into \hyperref[2]{Eq. (2)} leads to two linear homogeneous equations in A and B. These equations yield the following eigenfrequencies \(\Omega\), 
\begin{equation} \label{12}
    \begin{aligned}
        \left(\frac{\Omega}{\mu_0 \gamma}\right)^2 & = \big( M_0Dk^2 + H_0 + M_0f(k) \big) \\
        &\times \big( M_0Dk^2 + H_0 + M_0\frac{k_y^2}{k^2}(1-f(k)) \big).
    \end{aligned}
\end{equation}

\hyperref[12]{Eq. (12)} is widely used \cite{w1,w2,w3} to explain experimental data in thin layers, and is easily adapted to describe spin excitations in thin confined nanostructures (dots, stripes) in planar geometries. The magnetic potential outside the layer for \(x>a/2\) reads
\begin{equation} \label{13}
    \Phi^{+}_{out} = \sinh(\frac{ak}{2})\frac{e^{-kx}}{k}e^{i\Vec{k}\cdot\Vec{\rho}}(A-iB\frac{k_y}{k}).
\end{equation}

\noindent
and for \(x<-a/2\)      
\begin{equation}\label{14}
    \Phi^{-}_{out} = -\sinh(\frac{ak}{2})\frac{e^{-kx}}{k}e^{i\Vec{k}\cdot\Vec{\rho}}(A+iB\frac{k_y}{k}).
\end{equation}

We use \hyperref[13]{Eqs. (13)} and \hyperref[14]{(14)} to calculate the interlayer dipole interaction in a bilayer below.

\section{Bilayers}

\begin{figure} [htp]
\includegraphics[clip,width=0.9\columnwidth]{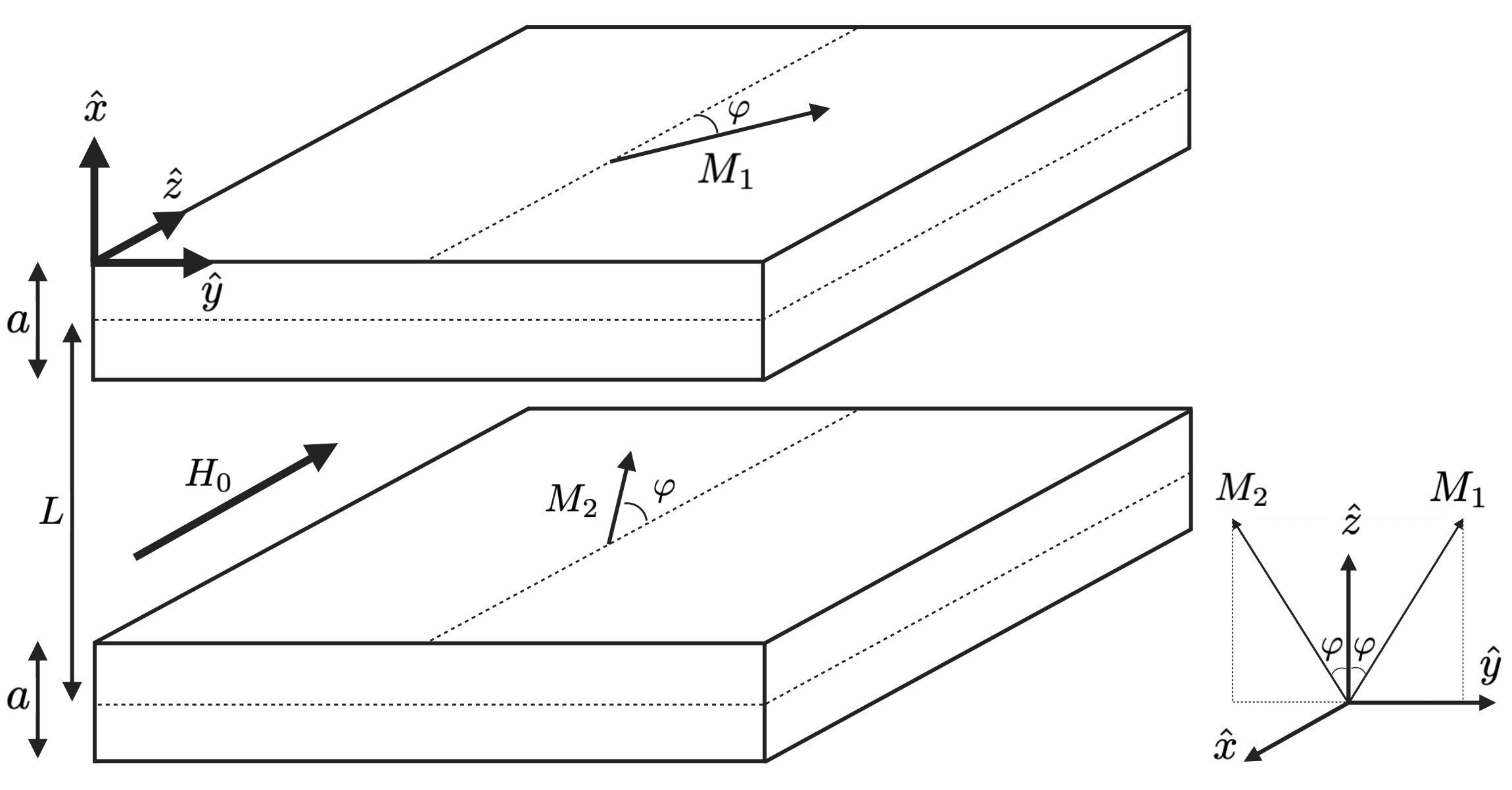}\label{2a}%

\caption{ \small Bilayer with the in-plane magnetization and antiferromagnetic interlayer exchange coupling.}
\label{fig2}
\end{figure}

Here we turn to the dipole-exchange SWs in antiferromagneticaly coupled magnetic bilayers, as shown in \hyperref[2a]{Fig. 2}.

Two FM layers with equal saturation magnetization \(M_0\) and exchange length \(D\) are coupled by an antiferromagnetic (AFM) exchange constant \(J>0\). Layer 1 is located at coordinates \((L-a)/2<x<(L+a)/2\), and the layer 2 at \(-(L+a)/2<x<-(L-a)/2\). The external magnetic field is applied along the z axis. The competition between the antiferromagnetic exchange interaction and the Zeeman energy leads to the formation of a canted magnetic structure, where the magnetization vectors deviate from the \(z\)-axis at field strengths \(0\leq H_0\leq2JM_0\) by an angle \(\varphi\),
\begin{equation}\label{15}
    \varphi = \cos^{-1}\left(\frac{H_0}{2JM_0}\right).
\end{equation}

\noindent
In this configuration the ground states for the first and second layer are \(M_1=(   0,M_0\sin\varphi,M_0 \cos\varphi)\) and \(M_2 = (0,-M_0 \sin\varphi,M_0 \cos\varphi)\), respectively.

We first consider the out-of-phase excitations where, as in the monolayer, we look for solutions in the form of plane waves:
\begin{equation}\label{16}
    \begin{aligned}
        \Vec{m}_1(\Vec{r})e^{i\Omega t} & = e^{i\Omega t}(-m_{1,x},-m_{1,y},m_{1,z}) \\
        & = e^{i\Omega t}(-A_1,-B_1\cos\varphi,B_1\sin\varphi) \\
        \Vec{m}_2(\Vec{r})e^{i\Omega t} & = e^{i\Omega t}(m_{2,x},m_{2,y},m_{2,z}) \\
        & = e^{i\Omega t}(A_2,B_2\cos\varphi,B_2\sin\varphi)
    \end{aligned}
\end{equation}

Using the same procedure as for the monolayer we calculate the magnetic potential outside layer 2 which leads to the dipolar field acting from layer 2 on layer 1, \(\Vec{h}_{21}\). After averaging over the thickness we get
\begin{equation}\label{17}
    \begin{aligned}
        h_{x,21} & = Ce^{i\Vec{k}\cdot\Vec{\rho}}\big(A_2-B_2\frac{ik_{2,\perp}}{k}\big), \\
        h_{y,21} & =-Ce^{i\Vec{k}\cdot\Vec{\rho}}\frac{ik_y}{k}\big(A_2-B_2\frac{ik_{2,\perp}}{k}\big), \\
        h_{z,21} & =-Ce^{i\Vec{k}\cdot\Vec{\rho}}\frac{ik_z}{k}\big(A_2-B_2\frac{ik_{2,\perp}}{k}\big), \\
    \end{aligned}
\end{equation}

\noindent
where \(C=4\frac{e^{-kL}\sinh^2(ak/2)}{ak}\), \(k_{2_\perp}=k_y\cos\varphi+k_z \sin\varphi\). In an analogous way we obtain for \(\Vec{h}_{12}\)
\begin{equation}\label{18}
    \begin{aligned}
        h_{x,12} & = -Ce^{i\Vec{k}\cdot\Vec{\rho}}\big(A_1+B_1\frac{ik_{1,\perp}}{k}\big), \\
        h_{y,12} & =-Ce^{i\Vec{k}\cdot\Vec{\rho}}\frac{ik_y}{k}\big(A_1+B_1\frac{ik_{1,\perp}}{k}\big), \\
        h_{z,12} & =-Ce^{i\Vec{k}\cdot\Vec{\rho}}\frac{ik_z}{k}\big(A_1+B_1\frac{ik_{1,\perp}}{k}\big), \\
    \end{aligned}
\end{equation}

\noindent
where \(k_{1_\perp}=k_y \cos\varphi-k_z \sin\varphi\). \hyperref[17]{Eqs. (17)} and \hyperref[18]{(18)} are substituted into the linearized LL equations for the magnetizations of the layers to find the frequency using the standard procedure. It is convenient to write these equations in matrix form. Collecting the terms for unknown constants A and B, and requiring the determinant to be zero we get
\begin{align} \label{19}
det\begin{bmatrix}
i\omega & -H_1 & ik_1 & T^+ \\
H & i\omega & T & -ik_2 \\
-ik_2  & T^+  & i\omega & -H_2  \\
T & ik_1 & H & i\omega
\end{bmatrix}=0,
\end{align}

\noindent
where \(\omega = \frac{\Omega}{\gamma \hbar}\) and
\begin{equation} \label{20}
    \begin{aligned}
    \centering
        H_i & = DM_0k^2 + H_0 \cos\varphi + M_0 \frac{k^2_{i,\perp}}{k^2}(1-f(k)) \\
        & - JM_0\cos(2\varphi), \\
        H & = DM_0k^2 + JM_0 + M_0f, \\
        k_i & = \frac{k_{i,\perp}}{|k|}M_0C,\\
        T & = -JM_0 + M_0C, \\
        T^+ & = JM_0\cos(2\varphi) + M_0C\frac{k_{1,\perp}}{|k|}\frac{k_{2,\perp}}{|k|}, \\
    \end{aligned}
\end{equation}

\noindent
where \(i=1,2\). The matrix is Hermitian so that we obtain a fourth-order equation for real frequencies as
\begin{equation} \label{21}
    \begin{aligned}
    \centering
        \omega^4 & + \Big[ 2T^+T - H_1H - H_2H + 2k_1k_2 \Big]\omega^2 \\
        & + 2\Big[ T(H_2k_1 - H_1k_2) + T^+(-k_1+k_2)H \Big]\omega\\
        & + \Big[ -H_1H_2T^2 - H_1Hk_2^2 + H_1H_2H^2 +(k_1k_2)^2 \\
        & -2T^+Tk_1k_2 +(T^+)^2T^2 - H_2Hk_1^2 - H^2(T^+)^2 \Big]\\
        & = 0.
    \end{aligned}
\end{equation}

\noindent
Doing the same derivation for in-phase excitations of the form
\begin{equation}\label{22}
    \begin{aligned}
        \Vec{m}_1(\Vec{r})e^{i\Omega t} & = e^{i\Omega t}(-m_{1,x},m_{1,y},-m_{1,z}) \\
        \Vec{m}_2(\Vec{r})e^{i\Omega t} & = e^{i\Omega t}(m_{2,x},m_{2,y},m_{2,z})
    \end{aligned}
\end{equation}

\noindent
we can prove that \hyperref[21]{Eq. (21)} turns out to be the same. Note that \hyperref[21]{Eq. (21)} is not invariant with respect to the change of direction of the wave vector. This might lead to non-reciprocal behavior, $\omega (k) \ne \omega (-k)$, as is common for dipolar interactions \cite{Tao,taylor}. Below, we discuss this non-reciprocity.

The dipole interaction between the layers disappears when \(k=0\), such that only the intralayer dipole interaction affects the frequency of the resonant mode since \(f\big|_{ak=0}= 1\) \cite{Teuling}. The interlayer dipole interaction is proportional to \(C=4\frac{e^{-kL}\sinh^2(ak/2)}{ak}\), which increases with the thickness of the layers and decreases exponentially with the distance between them, and may be disregarded when \(L \gg a\) and intralayer exchange dominates \cite{Teuling}. In the limit of \(L=a\), the interlayer dipole interactions may cause large nonreciprocities in synthetic antiferromagnets \cite{15}. In Ref. \cite{15}, giant nonreciprocal frequency shifts of propagating spin waves in interlayer exchange–coupled synthetic antiferromagnets is shown. This phenomenon is attributed to dipolar interactions between two magnetic layers in the bilayer. Furthermore, the authors of Ref. \cite{15} found that the sign of the frequency shift depends on relative configuration of the magnetizations in the bilayer.

The coefficient in the linear term of \hyperref[21]{Eq. (21)},
\begin{equation} \label{23}
    N_{nr} = 2\big[ T(H_2k_1 - H_1k_2) + T^+(-k_1+k_2)H \big],
\end{equation}

\noindent
changes sign when the sign of the wave vector flips, while the other coefficients in \hyperref[21]{Eq. (21)} remain invariant. Thus, \(N_{nr}\) is the only source of nonreciprocity.

Nonreciprocity in the dispersion is not always manifest, however. For example, in the collinear case (\(M_1=M_2, \varphi=0\)) at magnetic fields \(H_0 \geq 2JM_0\) the bilayer behaves like a ferromagnetic film and does not exhibit nonreciprocity, much like a spin valve with ferromagnetic coupling \cite{23}, as is apparent from \hyperref[20]{Eqs. (20)}-\hyperref[23]{(23)}.

In the general case of canted geometry, if the SW propagates perpendicular to the external field (\(k_z=0\)), we have \(k_{1,\perp}=k_{2,\perp}=k_y \cos\varphi\), \(H_1=H_2\), \(k_1=k_2\). Evidently, in such a case \(N_{nr}=0\) and the spectrum is reciprocal. If the SW instead propagates parallel to the external field (\(k_y=0\)) we have \(k_{2,\perp}=-k_{1,\perp}=k_z \sin\varphi\), \(H_1=H_2\), \(k_1=-k_2\), and \(N_{nr}\) explicitly depends on the direction of motion of the SW along or opposite to the field via 
\begin{equation} \label{24}
    N_{nr}=\mathrm{sign}(k_z)4M_0 C \sin\varphi[-TH_1+T^+ H_3 ].
\end{equation}

\noindent
\hyperref[fig3]{Fig. 3} shows the dispersion relation of the bilayer with the parameters for permalloy/Ru/permalloy. We see that the splitting of the bands along \((0,0,k_z)\) is larger than along \((0,0,-k_z)\) due to sign\((k_z)\) in \(N_{nr}\). 
\vspace{5mm}

\begin{figure} [htp]
\includegraphics[clip,width=0.9\columnwidth]{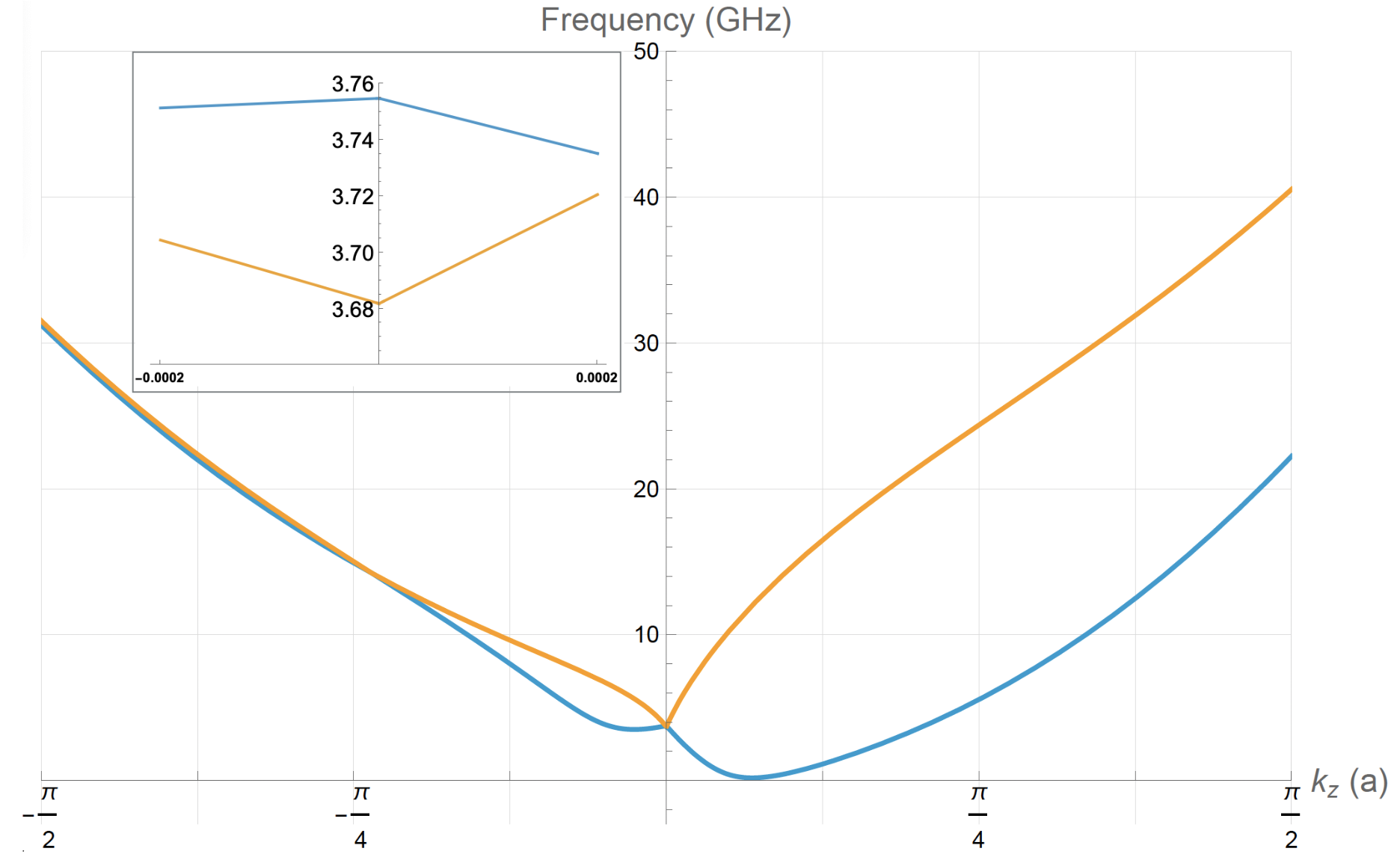}\label{fig3}%

\caption{ \small Dispersion relation of the bilayer for \(\varphi = \frac{\pi}{4}\), \(a = 7\) nm, \(L =10\) nm, \( J = 0.02\), \(D = 32.48\) nm$^2$ and \(M_0 = 7\cdot10^5\) A/m \cite{Belmeguenai}. The blue and orange curves are the acoustic and optical branches, respectively. The inset shows the gap around $k=0$.}
\label{fig3}
\end{figure}

\noindent
The gap between the bands at \(k=0\) for \(\varphi = \frac{\pi}{4} \) is
\begin{equation} \label{25}
    \frac{\Delta\omega}{\gamma \mu_0} =  \sqrt{\frac{H_0}{\sqrt{2}}} \left( \sqrt{\sqrt{2}H_0 + M_0} - \sqrt{\sqrt{2}H_0 + M_0(1-2J)}\right).
\end{equation}

A signature of spin waves is their microwave magnetic field (stray field), observable by NV center microscopy \cite{exp1}. For the  ferromagnetic film, \hyperref[1]{Eqs. (1)} and \hyperref[13]{(13)} are resonance frequencies  and magnetic potentials at \( x> \frac{a}{2} \), respectively. The  amplitudes A and B depend on the specific geometry of the sample and the power of the exciting microwaves. For general purpose, we can compute the coefficient normalized for a single magnon excitation \cite{exp2}. The time-indepent component \(m_z\) along the magnetization direction reads
\begin{equation} \label{26}
    m_z = -\frac{|m_x|^2 + |m_y|^2}{2M_0} = -\frac{|A|^2 + |B|^2}{2M_0} 
\end{equation}

\noindent
while according to the Hellmann-Feynman theorem for a single magnon
\begin{equation} \label{27}
    \frac{m_z}{M_0} = - \frac{\hbar}{\mu_B} \frac{\partial \Omega}{\partial \mu_0 H_0}
\end{equation}

\noindent
where \(\hbar\) is the reduced Planck constant and \(\mu_B\) the Bohr magneton. Substituting \(\Omega\) from \hyperref[12]{Eq. (12)} shows
\begin{equation}
\begin{aligned} \label{28}
    & |A|^2 + |B|^2 = \\
    & \frac{2M_0^2 \frac{\hbar}{\gamma \mu_B}\left(2Dk^2 + f(k) + \frac{k_y^2}{k^2}(1-f(k))\right)}{\sqrt{\Big(M_0Dk^2 + H_0 + M_0f(k)\Big)\left( M_0Dk^2 + H_0 + \frac{k_y^2}{k^2}M_0(1-f(k)) \right)}}
\end{aligned}
\end{equation}

\noindent
The ellipticity \(A/B\) follows from \hyperref[2]{Eq. (2)} as 
\begin{widetext}
\begin{equation} \label{29}
    \sqrt{\left(M_0Dk^2 + H_0 + M_0f(k)\right)\left( M_0Dk^2 + H_0 + \frac{k_y^2}{k^2}M_0(1-f(k)) \right)}A
    = -i\left(M_0Dk^2 +H_0 + M_0\frac{k_y^2}{k^2}(1-f(k))  \right)B.
\end{equation}
\end{widetext}

\noindent
\hyperref[28]{Eqs. (28)} and \hyperref[29]{(29)} determine the coefficients \(A\) and \(B\), and from \hyperref[13]{Eq. (13)} the stray field of a single magnon above the film (\(x>a/2\)) reads
\begin{equation} \label{30}
    h_{out}^+ = -M_0 Re\left\{\nabla \sinh(\frac{ka}{2})\frac{e^{-kx}}{k} e^{i\Vec{k}\cdot\Vec{\rho}} \left( A-B\frac{ik_y}{k} \right)\right\}.
\end{equation}

The situation is more complicated for a bilayer with non-collinear magnetizations with an “orbital correction” to the Hellmann-Feynman theorem \cite{exp3}. So we focus here on collinear structures, i.e. FM or AFM (\(H_0=0\)). However, the AFM case is not convenient to apply the Hellmann-Feymann theorem to as the external field is equal to zero.  
Consider the FM phase (\(H_0\geq2JM_0\)). The linear component in \hyperref[21]{Eq. (21)} vanishes, and the magnon dispersion reads
\begin{equation} \label{31}
\begin{aligned}
    & \frac{\Omega_{\pm}}{\gamma \mu_0} = \\
    & \sqrt{H_1H - (k_1^2+T^+T) \pm \sqrt{4(k_1^2T^+T) + (H_1T - HT^+)^2}},
\end{aligned}
\end{equation}

\noindent
where the functions are taken from \hyperref[20]{Eq. (20)} for \(\varphi=0\) (\(k_{1,\perp} = k_{2,\perp}\)). Both layers now contribute to the stray field so that

\begin{widetext}
\begin{equation} \label{32}
    h_{out,bilayer}^+ = -M_0 Re\left\{\nabla \sinh(\frac{ka}{2}) e^{i\Vec{k}\cdot\Vec{\rho}} \left[\frac{e^{-kx}}{k} \left( A_1-B_1\frac{ik_y}{k} \right) + \frac{e^{-k(L+x)}}{k}e^{i\Vec{k}\cdot\Vec{\rho}} \left( A_2-B_2\frac{ik_y}{k} \right) \right]\right\}.
\end{equation}
\end{widetext}

\noindent
where according to the Hellmann-Feynman theorem 

\begin{equation} \label{33}
    \frac{|A_1|^2 + |B_1|^2 + |A_2|^2 + |B_2|^2}{2M_0^2} = \frac{\hbar}{\mu_B} \frac{\partial \Omega}{\partial \mu_0 H_0},
\end{equation}

\noindent
which together with \hyperref[19]{Eqs. (19)} and \hyperref[20]{(20)} fully determines the problem.

\section{Conclusion}

In summary, this work demonstrates how the phenomenological framework developed by Bar'yakhtar and his collaborators addresses contemporary challenges in magnetism by revealing experimentally observed physical effects. We have analyzed the dipolar-exchange SW spectrum in thin ferromagnetic bilayers with an in-plane magnetization, incorporating AFM interlayer exchange and dipolar interactions within and between the layers. We highlight the effect of and requirements for the formation of nonreciprocal spin-wave propagation. The presented framework can be readily extended to include magnetocrystalline anisotropy, which influences the determination of the ground-state magnetic configuration but does not alter the calculation of dynamic dipole fields.

We now turn to the approximations employed in this study. Although the problem is solved exactly, this solution is valid only within the scope of the continuum approximation. It is well-suited for layers with thicknesses on the order of (tens of) nanometers. For extremely thin layers, such as atomically thin monolayers, two immediate issues arise. First, precise determination of the thickness \(a\) becomes challenging due to a generally non-flat atomic structure. We can, however, define the thickness as the vertical distance between similar magnetic atoms, as in Ref. \cite{Teuling}, so that our results are still valid by order of magnitude. Second, the continuum approach neglects the atomic-scale details of the layer's structure. This approximation is valid if the magnetic atoms of the monolayer form a square or hexagonal (triangular) lattice. For lattices with rectangular symmetry, especially those strongly elongated along one axis, additional dipole anisotropy may arise, potentially influencing the calculation of SW frequencies.

\begin{acknowledgments}
This publication is part of the project "Ronde Open Competitie XL" (file number OCENW.XL21.XL21.058) and "Ronde Open Competitie ENW pakket 21-3" (file number OCENW.M.21.215) which are (partly) financed by the Dutch Research Council (NWO). A.V.B. was supported by the EIC Pathfinder PALANTIRI project. E.V.T. was supported by the National Science Center of Poland, project no. UMO-2023/49/B/ST3/02920. G.B. was supported by JSPS Kakenhi Grants 22H04965 and JP24H02231. Images were made with BioRender.
\end{acknowledgments}

\bibliography{apssamp}% Produces the bibliography via BibTeX.

\end{document}